\documentclass[onecolumn,amsmath,amssymb,superscriptaddress,nofootinbib,11pt,a4paper]{revtex4}

\pdfoutput=1
\usepackage[T1]{fontenc}
\usepackage{CJK}
\usepackage{xcolor}
\usepackage{amsfonts}
\usepackage{epstopdf}
\usepackage{wrapfig} 
\usepackage{subcaption}
\usepackage{graphicx}  
\usepackage{dcolumn}   
\usepackage{bm}
\usepackage{float}
\usepackage[T1]{fontenc}
\usepackage{CJK}
\usepackage{xcolor}
\usepackage{amsfonts}
\usepackage{epstopdf}
\usepackage{wrapfig} 
\usepackage{subcaption} 
\usepackage{graphicx}  
\usepackage{dcolumn}   
\usepackage{bm}
\usepackage{float}
\usepackage[utf8]{inputenc}
\usepackage[T1]{fontenc}

\begin{document}

\title{Extracting Energy from Magnetized Rotating Black Holes in Horndeski Gravity via the Magnetic Penrose Process}
\author{Ke Wang}
\email{kkwwang2025@163.com}
\affiliation{School of Material Science and Engineering, Chongqing Jiaotong University, \\Chongqing 400074, China}

\author{Xiao-Xiong Zeng\footnote{Electronic address: xxzengphysics@163.com.  (Corresponding author)}}
\affiliation{College of Physics and Optoelectronic Engineering, Chongqing Normal University, \\Chongqing 401331, China}
\begin{abstract}
{ In Horndeski gravity, we investigate how to extract energy from a rotating black hole immersed in a uniform magnetic field $B$ based on the Magnetic Penrose Process. We map the ergosphere and negative energy regions of this spacetime, and analyze the relationship between the energy extraction efficiency and the hair parameter through both theoretical analysis and numerical simulations. The results show that the larger the hair parameter $h$, the smaller the ergosphere and negative energy regions of the black hole. For the same decay radius, in the case of $\hat{q} B \geq 0$, if the decay radius $r_x > 2$, the efficiency decreases as $h$ increases; if $r_x < 2$, the efficiency increases as $h$ increases; if $r_x = 2$, the efficiency is independent of $h$. However, when $\hat{q}B < 0$, except for the special case $r_x = 2$ where the efficiency is independent of $h$, the variation of efficiency with $h$ depends on the specific values of $r_x$ and $\hat{q}B$, and may exhibit either monotonic decrease or an initial increase followed by a decrease. We also find that in the absence of a magnetic field, the efficiency is negative and meaningless when $r_x > 2$, and such cases are excluded. In addition, when $\hat{q} B \geq 0$, the larger the $h$, the lower the maximum efficiency; when $\hat{q} B < 0$, in the case of a small magnetic field, the efficiency is negative and meaningless, while in the case of a large magnetic field, the efficiency of the black hole with hair is positive at high decay radius and reaches a high value, whereas the efficiency of the Kerr black hole remains negative.

}
\end{abstract}

\maketitle
 \newpage
\section{Introduction}
Extracting energy from rotating black holes has been a fascinating topic since it was proposed by Penrose. In the Penrose process \cite{10}, a particle splits into two parts inside the ergosphere, with the fragment possessing negative energy falling into the black hole, and the fragment with positive energy escaping to infinity, gaining more energy than the original particle, thereby extracting the rotational energy of the black hole. For an extreme Kerr black hole, the maximum efficiency is about $20.7\%$ \cite{8}, but for moderate spins (e.g., $a \leq 0.5M$) it drops to $2\%$, indicating that the smaller the spin, the lower the efficiency. In addition, since the fragments produced by the splitting require relative velocities exceeding half the speed of light \cite{11,12}, such events are extremely rare in astrophysics.

In the 1980s, with advancements in the study of matter electromagnetic interactions, research on the Penrose process revived, leading to the development of the magnetic Penrose process \cite{13,14,6,15,16,17}. An external magnetic field helps overcome the relative velocity constraint, as the energy of particles on negative energy orbits originates from matter electromagnetic interactions, thus no longer being subject to mechanical velocity conditions. The magnetic Penrose process can achieve extremely high energy extraction efficiencies, even easily exceeding 100$\%$. For example, around stellar-mass black holes, electrons can achieve efficiencies above 100$\%$ with magnetic fields as weak as milligauss \cite{18}. Unlike the mechanical Penrose process, the magnetic Penrose process does not rely on extremely high black hole spins and can operate efficiently even under low spin conditions. Moreover, magnetic fields of various strengths are widely distributed in the universe, ranging from $10^{-4}$ Gauss in the Galactic center region \cite{19} to $10^{16}$ Gauss on the surface of some magnetars \cite{20}. Many massive compact objects, such as Sagittarius A* and M87, are typically immersed in surrounding magnetic fields \cite{20,21,22,23,24,25,26,27}. In addition, in astrophysics, the magnetic Penrose process has been used to explain the origin and generation mechanism of extragalactic ultra high energy cosmic rays with energies exceeding $10^{20} eV$ \cite{28,29}, which lie beyond the GZK cutoff \cite{30,31,32,33}. Meanwhile, the magnetic Penrose process has also been applied to the study of relativistic jets \cite{34}. Therefore, research on the magnetic Penrose process is important and necessary. Currently, there have been some preliminary studies on the magnetic Penrose process \cite{35,36,37,38,39,40,41,5,42}.

General relativity, as the standard theory of classical gravity, has achieved great success on solar system scales and in gravitational wave observations. However, from a theoretical perspective, general relativity suffers from non-renormalizability in the ultraviolet regime, indicating that it needs to be replaced by a ultraviolet complete quantum theory of gravity at extreme energy scales. From an observational perspective, the late time accelerated expansion of the universe  and the inflationary mechanism in the early universe have also motivated explorations into the possibility of modifying general relativity. Among various modified gravity theories, Horndeski gravity is more popular \cite{43}. As early as 1974, Horndeski constructed the most general scalar-tensor theory in four dimensional spacetime, with an action that includes a scalar field $\phi$ coupled to the metric tensor $g_{\mu\nu}$, and whose corresponding equations of motion contain at most second order time and space derivatives \cite{43}. This property is crucial because it strictly avoids the Ostrogradsky instability, i.e., the ghost degrees of freedom typically present in higher derivative theories, thereby ensuring the classical stability of the theory. Horndeski's theory did not receive widespread attention for a long time, but after 2000, with the proposal of "generalized Galileon theories", it was realized that such theories are precisely a formulation of Horndeski gravity within a specific framework \cite{44,45}. Today, Horndeski gravity is recognized as the most general single field scalar-tensor theory, with its Lagrangian constructed from four arbitrary functions $G_2, G_3, G_4, G_5$, encompassing numerous classic models, including standard Quintessence, K-essence, Brans-Dicke theory, as well as the self-screening Chameleon and Symmetron theories \cite{46}. In black hole physics, this theory predicts black hole solutions and allows black holes to possess "hair" \cite{47}, which, in the era of gravitational wave astronomy, especially with the binary black hole merger events detected by LIGO/Virgo, provides crucial theoretical grounds for testing the strong field dynamics of gravity. Currently, many relevant properties of Horndeski gravity have been extensively studied \cite{48,1,2,49,50,51,52,53,54}. In view of the above theoretical value and observational significance, we will utilize the magnetic Penrose process to study how to extract energy from magnetized rotating black holes in Horndeski gravity, and we focus on   analyzing  the influence of the hair parameter on energy extraction.

The remainder of this paper is organized as follows. In Section 2, we introduce the motion of charged particles and the ergosphere in a magnetized rotating black hole in Horndeski gravity. In Section 3, we perform energy extraction via the magnetic Penrose process. We conclude in Section 4.

\section{Motion of charged particles and the ergosphere in a magnetized rotating black hole in Horndeski gravity}
\begin{figure}[!h]
  \centering
    \includegraphics[width=0.45\linewidth]{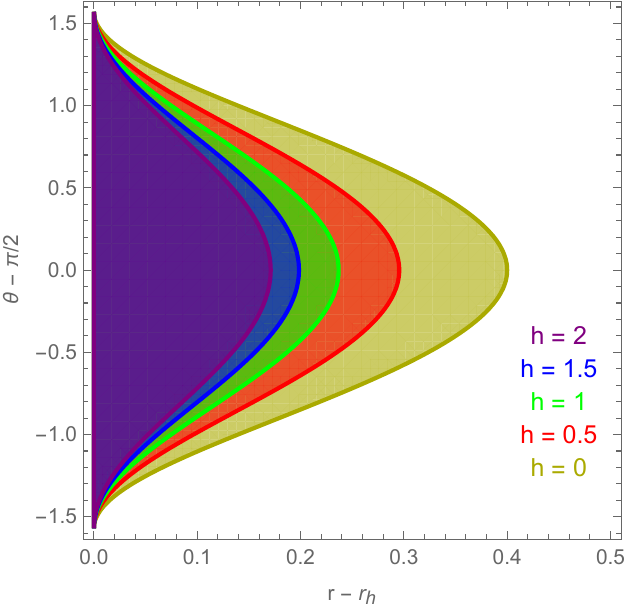} 
 \caption{The ergosphere region of the black hole for different values of $h$, with $a=0.8$}
\label{fig:1}
\end{figure}

In the Boyer-Lindquist coordinate system, the metric of a rotating black hole in Horndeski gravity is expressed as \cite{1,2}
\begin{equation}
\begin{aligned}
ds^2 = & -\left(\frac{\Delta - a^2\sin^2\theta}{\Sigma}\right)dt^2 + \frac{\Sigma}{\Delta} dr^2 + \Sigma d\theta^2 + \frac{2a\sin^2\theta}{\Sigma}\left(\Delta - (r^2 + a^2)\right)dtd\phi \\
& + \frac{\sin^2\theta}{\Sigma}\left[(r^2 + a^2)^2 - \Delta a^2\sin^2\theta\right]d\phi^2.
\end{aligned}
\end{equation}
The metric-related functions are
\begin{equation}
\Delta = r^2 + a^2 - 2Mr + hr \ln\left(\frac{r}{2M}\right),  \Sigma = r^2 + a^2\cos^2\theta.
\end{equation}
Here, $M$ is the black hole mass, $a$ is the black hole spin, and $h$ is the black hole hair. If $h=0$, the metric reduces to the Kerr metric. The event horizon of the black hole is determined by $\Delta=0$, which has two roots; the larger root is the event horizon, and the smaller root is the inner horizon. The boundary of the black hole ergosphere is located at $g_{tt}=0$. For simplicity, we set $M=1$. In Fig. \ref{fig:1}, we plot the ergosphere region of the black hole.
It can be seen from Fig. \ref{fig:1} that the larger $h$ is, the smaller the ergosphere becomes.

Now we immerse the black hole in an external uniform magnetic field. According to the Wald solution \cite{3}, the vector potential takes the form
\begin{equation}
A = aB\partial_t + \frac{B}{2}\partial_\phi.
\end{equation}
Adopting the gauge $A_{t}(r\rightarrow \infty) = 0$ \cite{4}, the vector potential is obtained as
\begin{equation}
A_{t} = aB(1 + g_{tt}) + \frac{B}{2} g_{t\phi}, \quad A_{\phi} = aBg_{t\phi} + \frac{B}{2} g_{\phi \phi}, \quad A_{r} = A_{\theta} = 0.
\end{equation}

\begin{figure}[!h]
  \centering
  \begin{subfigure}{0.35\textwidth}
    \centering
    \includegraphics[width=\linewidth]{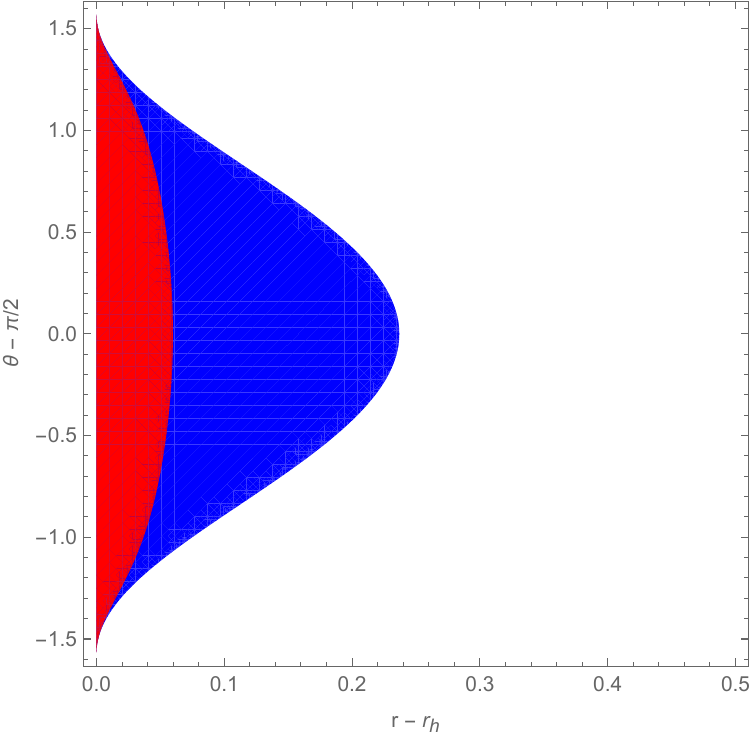} 
  \end{subfigure}
  \begin{subfigure}{0.35\textwidth}
    \centering
    \includegraphics[width=\linewidth]{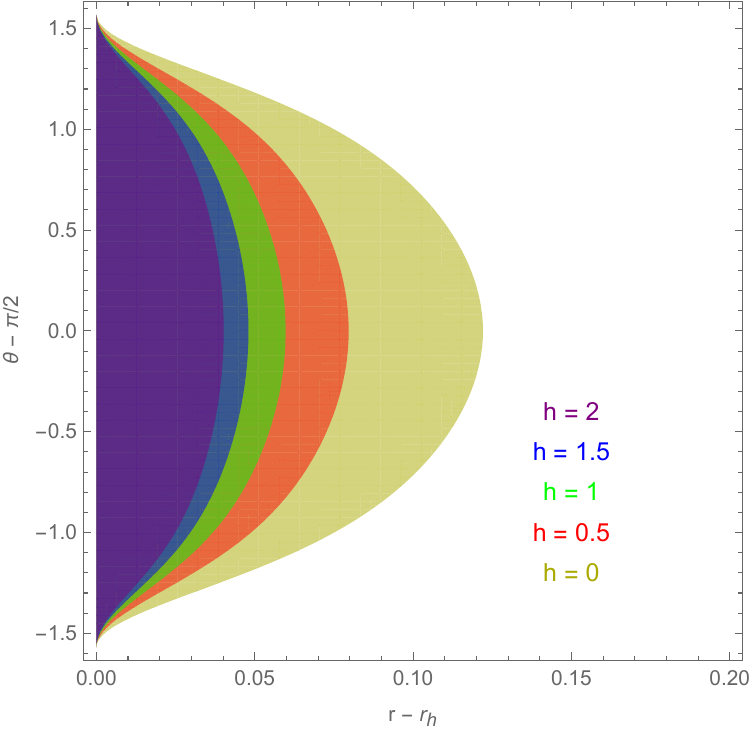} 
  \end{subfigure}
 \caption{$\bar{q}B=0$, $a=0.8$, $\bar{L}=-1$, left panel: $h=1$, red indicates the negative-energy region, blue indicates the ergosphere; right panel: the negative energy region of the particle for different values of $h$}
  \label{fig:2}
\end{figure}

\begin{figure}[!h]
  \centering
  \begin{subfigure}{0.35\textwidth}
    \centering
    \includegraphics[width=\linewidth]{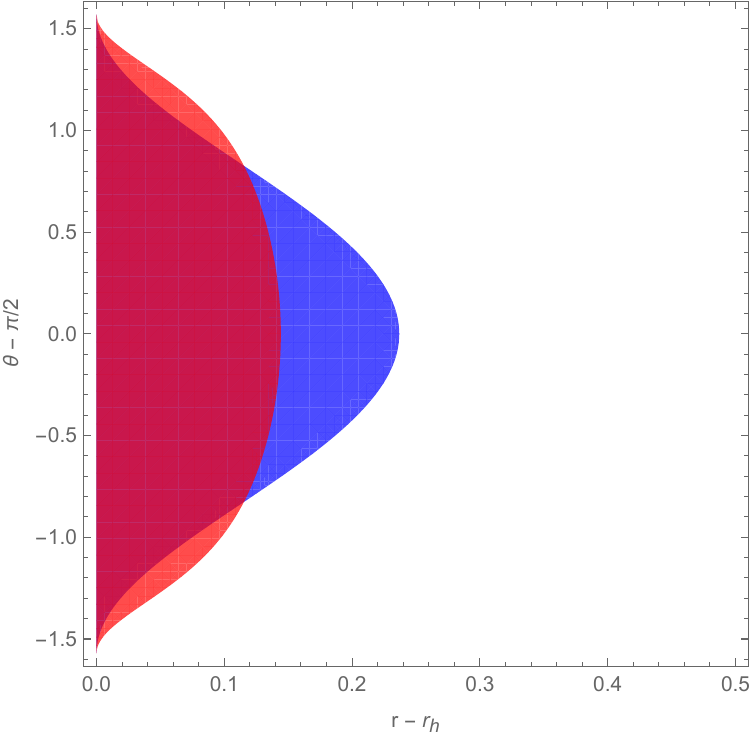} 
  \end{subfigure}
  \begin{subfigure}{0.35\textwidth}
    \centering
    \includegraphics[width=\linewidth]{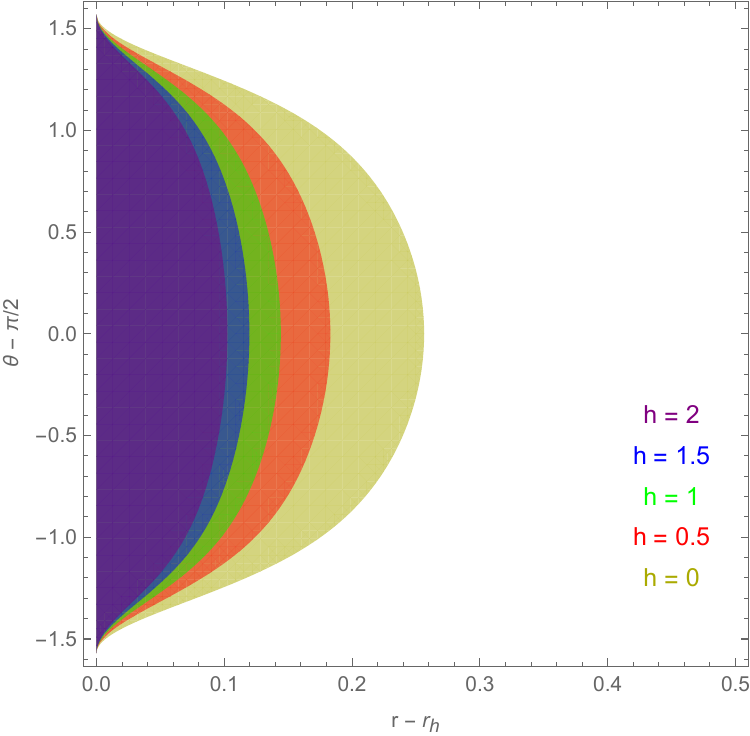} 
  \end{subfigure}
 \begin{subfigure}{0.35\textwidth}
    \centering
    \includegraphics[width=\linewidth]{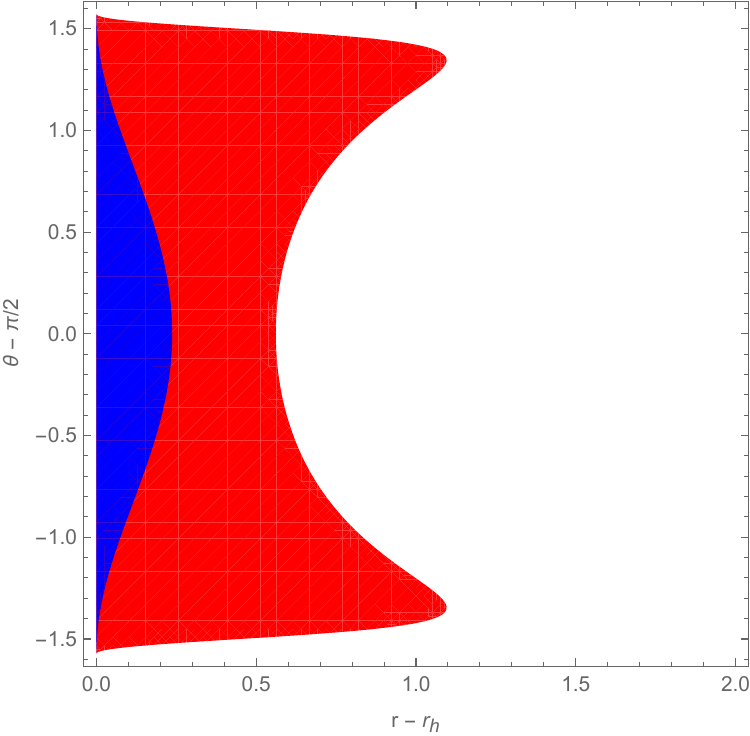}  
  \end{subfigure}
 \begin{subfigure}{0.35\textwidth}
    \centering
    \includegraphics[width=\linewidth]{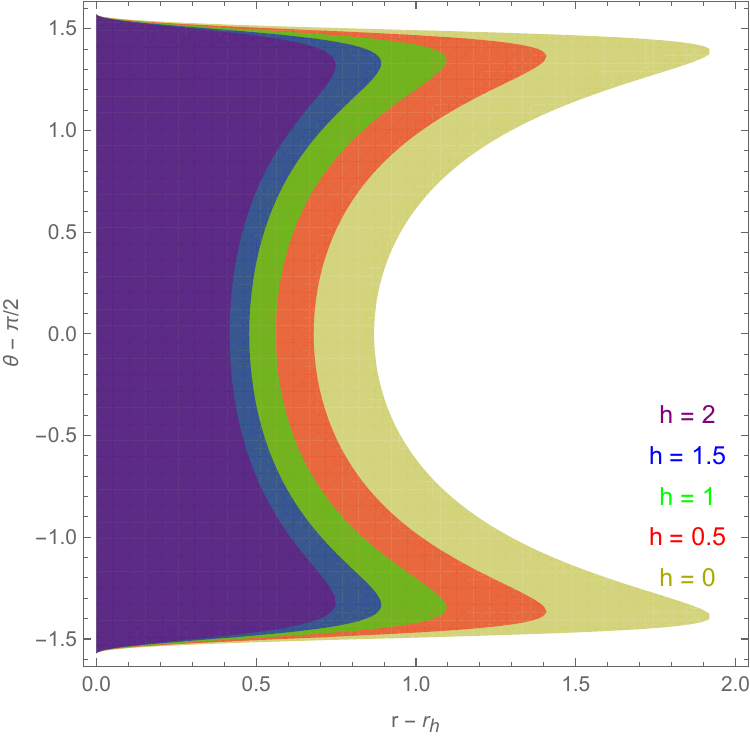} 
  \end{subfigure}
 \begin{subfigure}{0.35\textwidth}
    \centering
    \includegraphics[width=\linewidth]{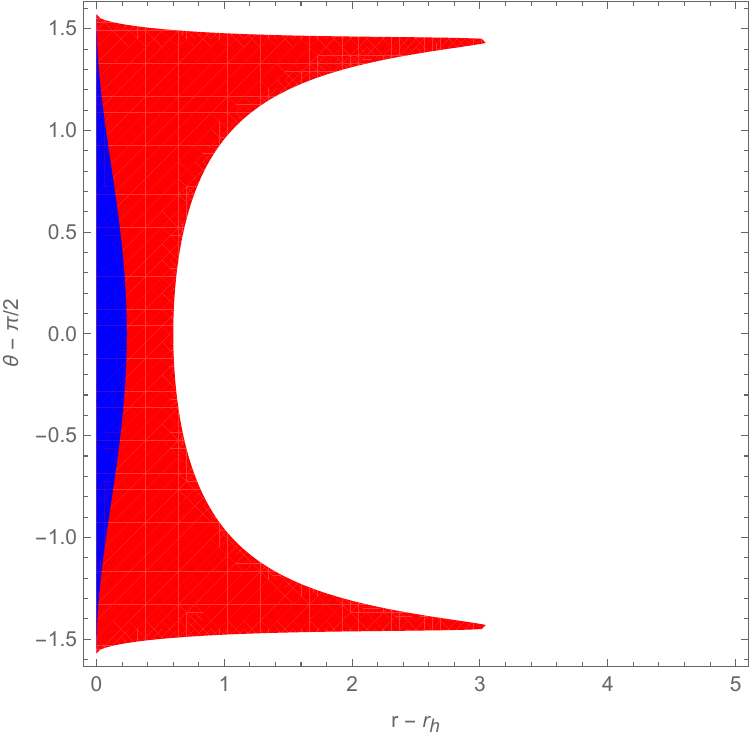}  
  \end{subfigure}
 \begin{subfigure}{0.35\textwidth}
    \centering
    \includegraphics[width=\linewidth]{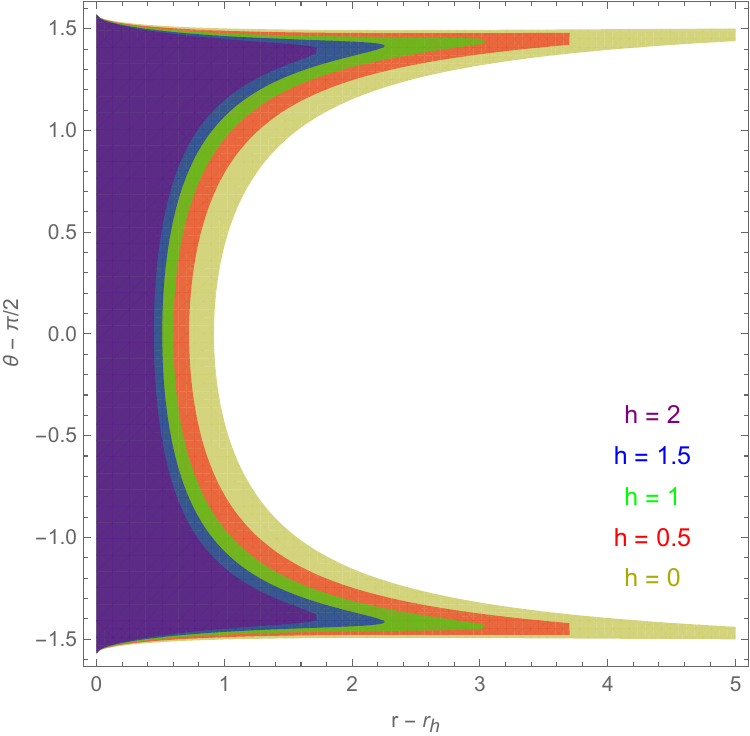} 
  \end{subfigure} 
  \caption{$\bar{q}B>0$, $a=0.8$, left panel: $h=1$, red indicates the negative energy region, blue indicates the ergosphere; right panel: the negative energy region of the particle for different values of $h$. Top row: $\bar{q}B=0.2$, middle and bottom rows: $\bar{q}B=5$, top and middle rows: $\bar{L}=-1$, bottom row: $\bar{L}=1$}
  \label{fig:3}
\end{figure}

\begin{figure}[!h]
  \centering
  \begin{subfigure}{0.35\textwidth}
    \centering
    \includegraphics[width=\linewidth]{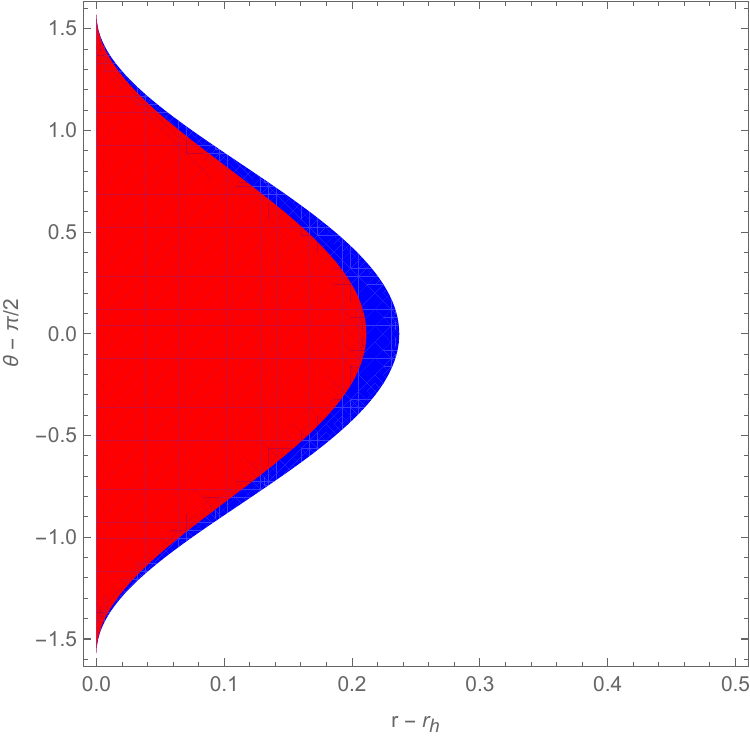} 
  \end{subfigure}
  \begin{subfigure}{0.35\textwidth}
    \centering
    \includegraphics[width=\linewidth]{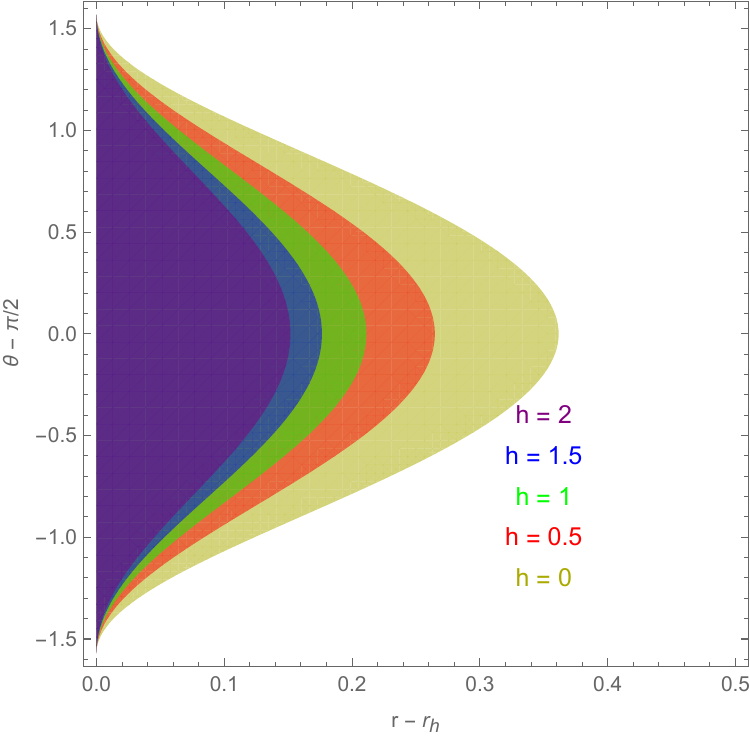} 
  \end{subfigure}
  \begin{subfigure}{0.35\textwidth}
    \centering
    \includegraphics[width=\linewidth]{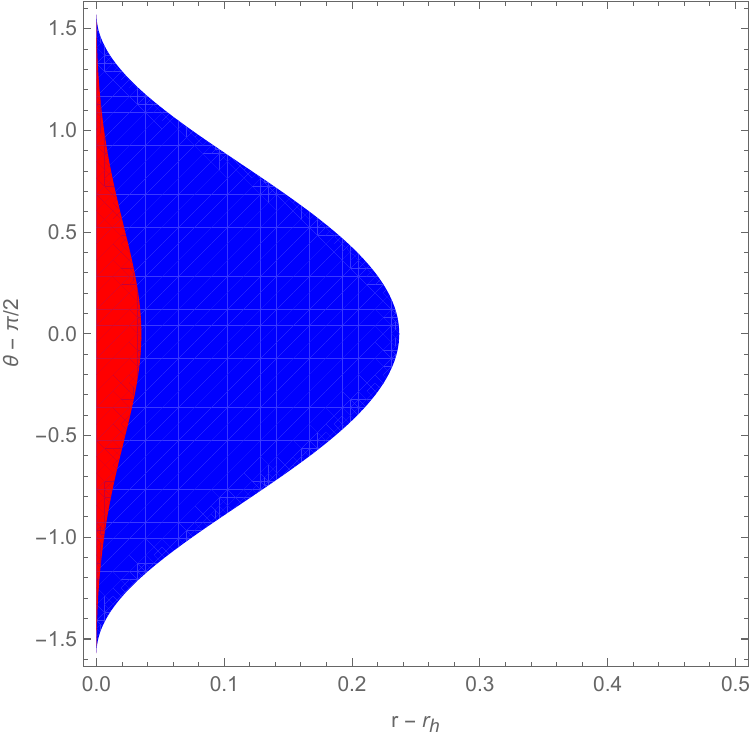} 
  \end{subfigure}
  \begin{subfigure}{0.35\textwidth}
    \centering
    \includegraphics[width=\linewidth]{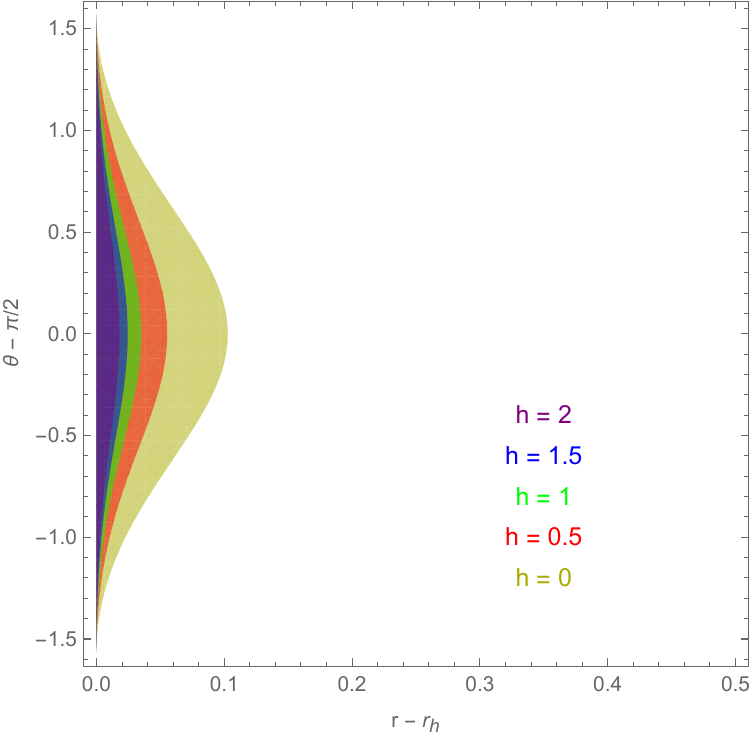} 
  \end{subfigure}
  \caption{$\bar{q}B<0.2$, $a=0.8$, $\bar{L}=-10$, left panel: $h=1$, red indicates the negative energy region, blue indicates the ergosphere; right panel: the negative energy region of the particle for different values of $h$. Top row: $\bar{q}B=-0.2$, bottom row: $\bar{q}B=-2$}
  \label{fig:4}
\end{figure}

The motion of a charged particle in this spacetime can be described by the Lagrangian
\begin{equation}
\mathcal{L}=\frac{1}{2}m\,g_{\mu\nu}\,\frac{dx^{\mu}}{d\tau}\frac{dx^{\nu}}{d\tau}+q\,A_{\mu}\,\frac{dx^{\mu}}{d\tau},
\end{equation}
where $m$, $q$, $\tau$ are the mass, charge, and proper time of the particle, respectively. Since the spacetime is axisymmetric, the Lagrangian does not explicitly depend on the two coordinates ${t,\phi}$. Therefore, their conjugate momenta are constants, i.e.,
\begin{equation}
\frac{\partial\mathcal{L}}{\partial\dot{t}} =mU_{t}+qA_{t}=-E,\frac{\partial\mathcal{L}}{\partial\dot{\phi}} =mU_{\phi}+qA_{\phi}= L,\label{6}
\end{equation}
where $U_{\mu}$ is the covariant 4-velocity of the particle, the dot denotes the derivative with respect to the proper time $\tau$, and $E$ and $L$ are the energy and angular momentum of the particle, respectively. Assuming the particle moves in the equatorial plane, its 4-velocity can be expressed as $U^{\mu} = \dot{t}(1, v, 0, \Omega)$, where $v = dr/dt$, $\Omega = d\phi/dt$. According to the normalization condition $U_{\mu}U^{\mu} = -1$, we obtain
\begin{equation}
\dot{t}^{2}\left(g_{tt}+g_{rr}v^{2}+g_{\phi\phi}\Omega^{2}+2g_{t \phi}\Omega\right)=-1.
\end{equation}
Solving the above equation yields
\begin{equation}
\Omega=-\frac{g_{t\phi}}{g_{\phi\phi}}\pm\sqrt{\left(\frac{g_{t\phi }}{g_{\phi\phi}}\right)^{2}-\frac{1}{g_{\phi\phi}}\left[g_{tt}+g_{rr}v^{2}+\frac {1}{(\dot{t})^{2}}\right]},\label{8}
\end{equation}
where $+$ and $-$ correspond to corotating and counter-rotating motions, respectively. For any particle moving in the equatorial plane, its angular velocity is constrained by
\begin{equation}
\Omega_{-}\leq\Omega\leq\Omega_{+},\Omega_{\pm}=-\frac{g_{t \phi}}{g_{\phi\phi}}\pm\sqrt{\left(\frac{g_{t\phi}}{g_{\phi\phi}}\right)^{2}- \frac{1}{g_{\phi\phi}}\left(g_{tt}+g_{rr}v^{2}\right)},
\end{equation}
where $\Omega_{\pm}$ correspond to the motion of photons. The effective potential of the particle can be expressed as \cite{5}
\begin{equation}
V_{\rm eff}(r)
=-\bar{q}A_{t}-\frac{g_{t\phi}}{g_{\phi\phi}}(\bar{L}-\bar{q}A_{\phi}) +\sqrt{-\frac{1}{g^{tt}}\left[\frac{(\bar{L}-\bar{q}A_{\phi})^{2}}{g_{\phi\phi}}+1 \right]},
\end{equation}
where $\bar{q}=q/m$, $\bar{L}=L/m$ denote the charge-to-mass ratio and the specific angular momentum of the particle, respectively. The negative energy region is defined as the region where $V_{\rm eff}(r)<0$, which is the region where energy can be extracted; the reasoning can be found in \cite{5}. We plot the negative energy region of the particle in Figs. \ref{fig:2}, \ref{fig:3}, and \ref{fig:4}. First, we consider the case without a magnetic field, i.e., $\bar{q}B=0$. In this case, the negative energy region exists only when $\bar{L}<0$.
It can be seen from Fig. \ref{fig:2} that in the absence of a magnetic field, the negative energy region lies inside the ergosphere, and the larger $h$ is, the smaller the negative energy region becomes, which is similar to the variation of the ergosphere.

Figure \ref{fig:3} shows the case of $\bar{q}B>0$. It can be seen that only a small $\bar{q}B$, e.g., $\bar{q}B=0.2$, is needed for the negative energy region to extend beyond the ergosphere at some angles. As $\bar{q}B$ increases, the negative-energy region expands continuously until it completely extends beyond the ergosphere. In addition, when $\bar{q}B$ is sufficiently large, as shown in the bottom row of Fig. \ref{fig:3}, the negative energy region also exists in the region of $\bar{L}>0$. Furthermore, similar to the case of $\bar{q}B=0$, the larger $h$ is, the smaller the negative energy region becomes.

Figure \ref{fig:4} shows the case of $\bar{q}B<0$. In this case, for the negative energy region to exist, it must be that $\bar{L}<0$, and the negative energy region lies inside the ergosphere. The more negative $\bar{q}B$ is, the smaller the negative energy region becomes, until it disappears. Similarly, the larger $h$ is, the smaller the negative-energy region becomes.

\section{Energy extraction via the magnetic Penrose process}
Now, we simplify the study of Figs. \ref{fig:2} to \ref{fig:4} by considering the particle motion confined to the equatorial plane. A particle 0 splits into particles 1 and 2. In this process, four momentum and charge are conserved, so we have
\begin{equation}
m_0U_0^\mu = m_1U_1^\mu + m_2U_2^\mu,q_0 = q_1 + q_2.
\end{equation}
The subscripts $0$, $1$, and $2$ represent particles $0$, $1$, and $2$, respectively. Combining the above relations, energy and angular momentum conservation give
\begin{equation}
E_0 = E_1 + E_2,L_0 = L_1 + L_2.
\end{equation}
The three components of four momentum conservation can be written as
\begin{equation}
{m_0\dot{t}_0 = m_1\dot{t}_1 + m_2\dot{t}_2,} {m_0v_0\dot{t}_0 = m_1v_1\dot{t}_1 + m_2v_2\dot{t}_2,} {m_0\Omega_0\dot{t}_0 = m_1\Omega_1\dot{t}_1 + m_2\Omega_2\dot{t}_2.} 
\end{equation}
From equation \eqref{6}, we obtain
\begin{equation}
E_{i} = -m_{i}\dot{t}_{i}X_{i} - q_{i}A_{t},  X_{i} =g_{tt} + \Omega_{i}g_{t\phi}.\label{14}
\end{equation}
Thus, the energy of particle 2 can be written as
\begin{equation}
E_{2} = \kappa (E_{0} + q_{0}A_{t}) - q_{2}A_{t},\kappa = \frac{\Omega_0 - \Omega_1 X_2}{\Omega_2 - \Omega_1 X_0} = \frac{v_0 X_1 - v_1 X_0 X_2}{v_2 X_1 - v_1 X_2 X_0}. \label{15}
\end{equation}
The energy extraction efficiency is defined as
\begin{equation}
\eta = \frac{E_2 - E_0}{E_0}.
\end{equation}
According to equation \eqref{15}, the efficiency can be expressed as
\begin{equation}
\eta = \kappa -1 + \kappa \frac{\bar{q}_0 A_t}{E_0/m_0} - \frac{m_2}{m_0} \frac{\bar{q}_2 A_t}{E_0/m_0}.
\end{equation}
Following the literature \cite{6}, we set $q_0=0$, so that $q_1=-q_2=q$. Considering that in almost all realistic scenarios the incident particle is initially non-relativistic \cite{6}, we adopt $E_0/m_0=1$. In this case, the angular velocity of particle 0 can be derived using \eqref{8} and \eqref{14}
\begin{equation}
\Omega_0 = \frac{-g_{t\phi}(1 + g_{tt}) + \sqrt{(g_{t\phi}^2 - g_{tt}g_{\phi\phi})(1 + g_{tt})}}{g_{\phi\phi} + g_{t\phi}^2}.
\end{equation}
We consider the maximum energy extraction efficiency. As shown in \cite{6}, $\eta$ is maximized when the radial velocities are zero and the angular velocities of the two fragments take their limiting values, i.e., $\Omega_{1} = \Omega_{-}$ and $\Omega_{2} = \Omega_{+}$. In this case, the efficiency becomes as
\begin{equation}
\eta = \frac{1}{2}\left[\sqrt{1 + g_{tt}} -1\right] + \hat{q} A_t, \label{19}
\end{equation}
where $\hat{q} = \frac{m_1}{m_0} \bar{q}$ is referred  as the rescaled charge-to-mass ratio of particle 1. Thus, under the above approximation, $\eta$ depends only on four parameters: ${a, h, \hat{q} B}$ and the splitting point $r = r_{x}$. In addition, the mass ratio $m_1 / m_0$ is subject to the constraint
\begin{equation}
m_1 + m_2 \leq m_0.
\end{equation}
Substituting the metric components and the vector potential into equation \eqref{19}, we obtain
\begin{equation}
\eta = \frac{1}{2} \left( \sqrt{\frac{2 - h \ln(r_{x}/2)}{r_{x}}} - 1 \right) + \frac{a \hat{q} B}{2} \cdot \frac{2 - h \ln(r_{x}/2)}{r_{x}}.\label{21}
\end{equation}
For a non-magnetic extreme Kerr black hole, taking $h=0$, $B=0$, $r_{x}=r_h$, $a=1$, and substituting into the above expression yields a maximum efficiency of $0.207$, which is consistent with known results \cite{8}. We now perform a theoretical analysis of the relationship between the efficiency expression \eqref{21} and $h$. $r_{x}$ must be greater than the event horizon, and the event horizon lies in the region $r<2$. Then, taking the partial derivative of equation \eqref{21} with respect to $h$, we obtain
\begin{equation}
\frac{\partial \eta}{\partial h} = -\frac{\ln(r_{x}/2)}{4r_{x} \sqrt{\frac{2 - h\ln(r_{x}/2)}{r_{x}}}} - \frac{a \hat{q} B \ln(r_{x}/2)}{2r_{x}}.
\end{equation}
It is easy to see that in the case $\hat{q} B \geq 0$, if $r_x > 2$, then $\partial \eta/\partial h < 0$, and the efficiency decreases as $h$ increases; if $r_x < 2$, then $\partial \eta/\partial h > 0$, and the efficiency increases as $h$ increases; if $r_x = 2$, then $\partial \eta/\partial h = 0$, and the efficiency is independent of $h$. When $\hat{q}B < 0$, let $c = -\hat{q}B > 0$; then the partial derivative becomes
\begin{equation}
\frac{\partial \eta}{\partial h} = \frac{\ln(r_x/2)}{r_x} \left( -\frac{1}{4\sqrt{\frac{2 - h\ln(r_x/2)}{r_x}}} + \frac{a c}{2} \right).
\end{equation}
Here $h, a, r_x > 0$, and it is required that the expression under the square root is non-negative ($2 - h\ln(r_x/2) \ge 0$). We now discuss the cases according to the relationship between $r_x$ and $2$. When $r_x = 2$, $\ln(r_x/2)=0$, so $\partial \eta/\partial h = 0$, and the efficiency is independent of $h$. When $r_x > 2$, then $\ln(r_x/2) > 0$, and the range of $h$ is $0 < h \le 2/\ln(r_x/2)$. Denote the bracket as
\begin{equation}
S(h) = -\frac{1}{4\sqrt{\frac{2 - h\ln(r_x/2)}{r_x}}} + \frac{a c}{2}.
\end{equation}
$S(h)$ is strictly decreasing with increasing $h$, and
\begin{equation}
S(0^+) = \frac{ac}{2} - \frac{1}{4\sqrt{2/r_x}}, \quad \lim_{h\to h_{\max}^-} S(h) = -\infty.
\end{equation}
If $S(0^+) > 0$, i.e., $ac > \frac{1}{2\sqrt{2/r_x}} = \frac{\sqrt{r_x}}{2\sqrt{2}}$, then there exists a unique $h_0\in(0,h_{\max})$ such that $S(h_0)=0$. When $h < h_0$, $S>0$, so $\partial\eta/\partial h>0$; when $h > h_0$, $S<0$, so $\partial\eta/\partial h<0$. The efficiency first increases and then decreases, reaching a maximum at $h_0$. If $S(0^+) \le 0$, i.e., $ac \le \frac{\sqrt{r_x}}{2\sqrt{2}}$, then $S(h)<0$ for all $h\in(0,h_{\max})$, so $\partial\eta/\partial h<0$, and the efficiency decreases monotonically as $h$ increases. When $r_x < 2$, then $\ln(r_x/2) < 0$, and $h$ can take any positive value (the expression under the square root is always positive). Denote $S(h)$ as above, but now $S(h)$ is strictly increasing with increasing $h$, and
\begin{equation}
S(0) = \frac{a c}{2} - \frac{1}{4\sqrt{2/r_x}}, \quad \lim_{h\to\infty} S(h) = \frac{a c}{2} > 0.
\end{equation}
Since $\ln(r_x/2)/r_x < 0$, the sign of $\partial \eta/\partial h$ is opposite to that of $S(h)$. If $S(0) \ge 0$, i.e., $a c \ge \frac{\sqrt{r_x}}{2\sqrt{2}}$, then $S(h) \ge 0$ holds for all $h$, so $\partial \eta/\partial h \le 0$ (the equality holds only when $S(0)=0$ and $h=0$; for $h>0$, it is strictly negative), and the efficiency decreases as $h$ increases. If $S(0) < 0$, i.e., $a c < \frac{\sqrt{r_x}}{2\sqrt{2}}$, then there exists a unique $h_0>0$ such that $S(h_0)=0$. When $h < h_0$, $S<0$, so $\partial \eta/\partial h >0$; when $h > h_0$, $S>0$, so $\partial \eta/\partial h <0$. The efficiency first increases and then decreases, reaching a maximum at $h_0$. Substituting $c = -\hat{q}B$, we obtain $a c = -a\hat{q}B$. Therefore, the above conditions can be rewritten as: when $r_x < 2$ and $-a\hat{q}B \ge \frac{\sqrt{r_x}}{2\sqrt{2}}$ (i.e., $\hat{q}B \le -\frac{\sqrt{r_x}}{2\sqrt{2}a}$), the efficiency decreases as $h$ increases; when $r_x < 2$ and $0 > \hat{q}B > -\frac{\sqrt{r_x}}{2\sqrt{2}a}$, the efficiency first increases and then decreases, with a maximum at a certain point. In summary, when $\hat{q}B<0$, except for the special case $r_x=2$, the variation of efficiency with $h$ depends on the specific values of $r_x$ and $\hat{q}B$, and may exhibit either monotonic decrease or an initial increase followed by a decrease.

Next, we analyze the influence of various parameters on the efficiency through numerical calculations based on equation \eqref{21}. First, we consider the case without a magnetic field.
\begin{figure}[!h]
  \centering
  \begin{subfigure}{0.45\textwidth}
    \centering
    \includegraphics[width=\linewidth]{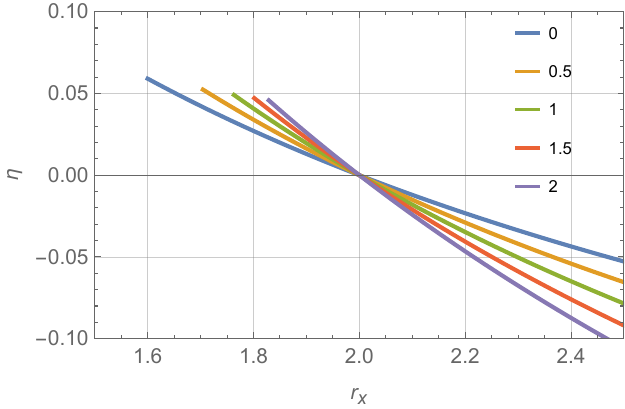} 
  \end{subfigure}
  \begin{subfigure}{0.45\textwidth}
    \centering
    \includegraphics[width=\linewidth]{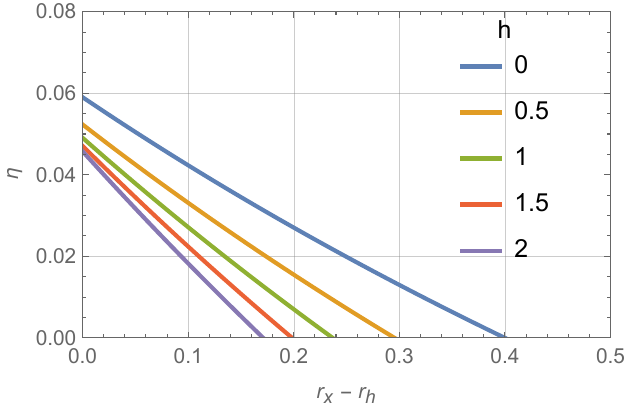} 
  \end{subfigure}
  \begin{subfigure}{0.45\textwidth}
    \centering
    \includegraphics[width=\linewidth]{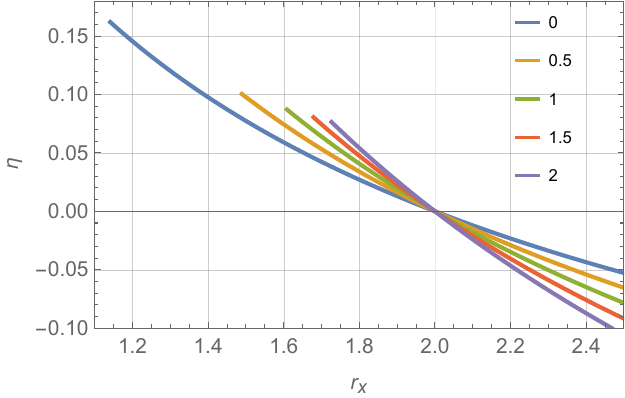} 
  \end{subfigure}
  \begin{subfigure}{0.45\textwidth}
    \centering
    \includegraphics[width=\linewidth]{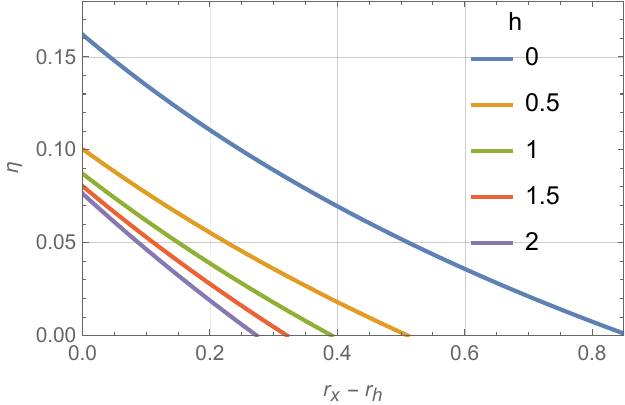} 
  \end{subfigure}
  \caption{$B=0$, the variation of energy extraction efficiency with the decay radius for different values of $h$. The left and right panels show different horizontal axis scales. Upper panel: $a=0.8$, lower panel: $a=0.99$.}
\label{fig:5}
\end{figure}
It can be seen from the left panel of Fig. \ref{fig:5} that for the same decay radius, when $r_x < 2$, the larger $h$ is, the higher the energy extraction efficiency; when $r_x > 2$, the larger $h$ is, the lower the energy extraction efficiency, which is consistent with the theoretical derivation. However, when $r_x > 2$, the efficiency is negative and meaningless, so such cases are excluded. Therefore, in the absence of a magnetic field, for the same decay radius, the larger $h$ is, the higher the efficiency. From the right panel of Fig. \ref{fig:5}, it can be seen that the farther the decay radius is from the event horizon, the lower the efficiency. The right panel also shows that the larger $h$ is, the lower the maximum efficiency. On the other hand, the higher the spin, the higher the efficiency.

\begin{figure}[!h]
  \centering
  \begin{subfigure}{0.45\textwidth}
    \centering
    \includegraphics[width=\linewidth]{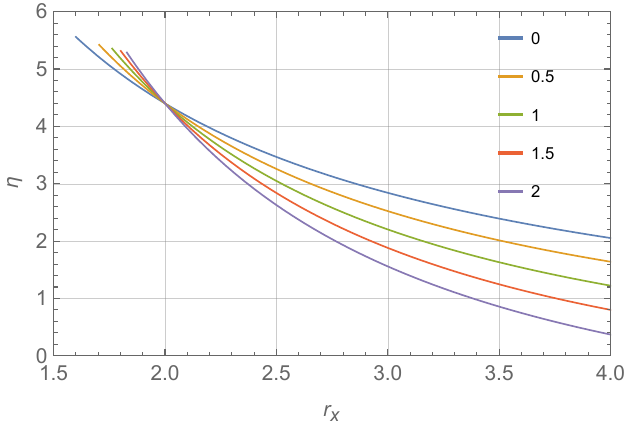} 
  \end{subfigure}
  \begin{subfigure}{0.45\textwidth}
    \centering
    \includegraphics[width=\linewidth]{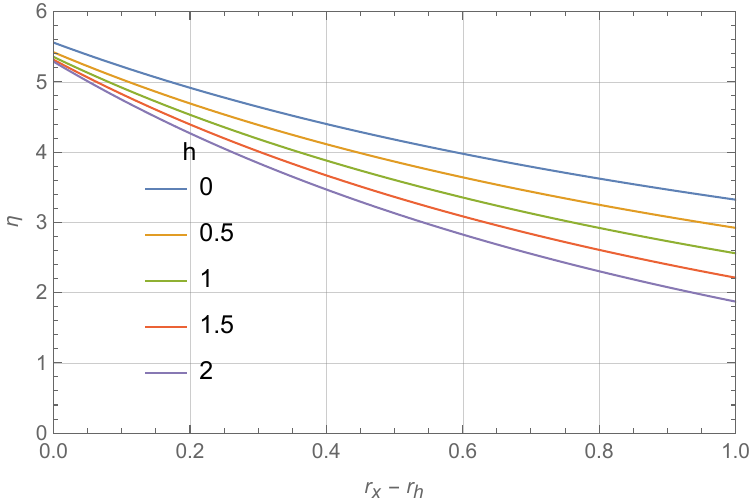} 
  \end{subfigure}
  \begin{subfigure}{0.45\textwidth}
    \centering
    \includegraphics[width=\linewidth]{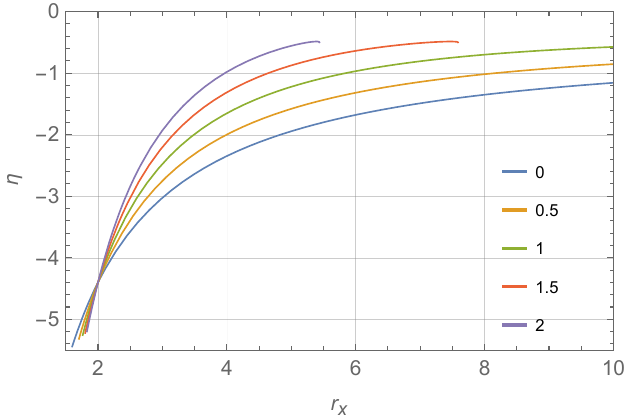} 
  \end{subfigure}
  \begin{subfigure}{0.45\textwidth}
    \centering
    \includegraphics[width=\linewidth]{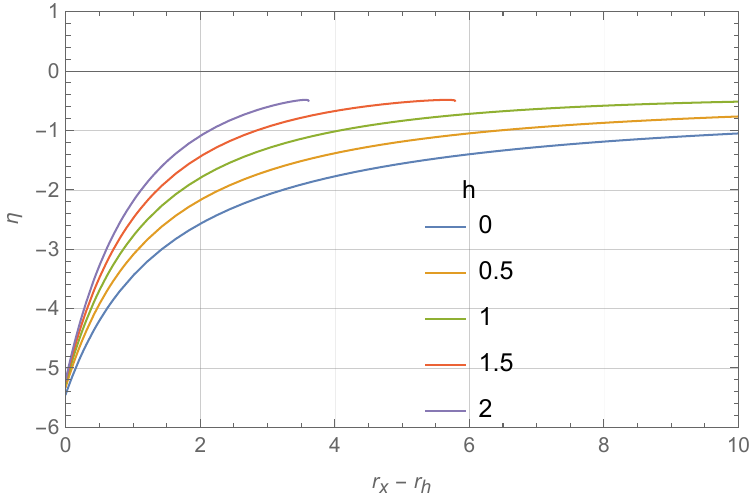} 
  \end{subfigure}
 \caption{$a=0.8$, the variation of energy extraction efficiency with the decay radius for different values of $h$. The left and right panels show different horizontal axis scales. Upper panel: $B=-10^{-17}$, lower panel: $B=10^{-17}$.}
\label{fig:6}
\end{figure}

When electromagnetic interactions are included, the situation changes. Without loss of generality, we follow \cite{7} to analyze a specific process: a neutron (particle 0) decays into an electron $e^{-}$ (particle 1), a proton $p^{+}$ (particle 2), and an antineutrino $\bar{\nu}{e}$, i.e.,
\begin{equation}
n \rightarrow e^{-} + p^{+} + \bar{\nu}{e}.
\end{equation}
Here, the effect of the antineutrino on the proton energy is neglected. In units where $C=G=\hbar=4\pi\varepsilon_0=1$, $m_{1} / m_{0} \sim 1 / 1839$ and $\bar{q} = q_{e} / m_{e} \sim - 2.04 \times 10^{21}$. The rescaled charge-to-mass ratio reaches an extremely high value $\hat{q} \sim - 1.1 \times 10^{18}$. For $|B| > 10^{-18}$, the efficiency is predominantly governed by the electromagnetic term. In Fig. \ref{fig:6}, we consider the case of $|B| = 10^{-17}$. In physical units, for $ M = M_{\odot}$, this corresponds to $|B| \sim 2.35 \times 10^{2}$ Gauss; for $M = 10^{6} M_{\odot}$, $|B| \sim 2.35 \times 10^{-4}$ Gauss; for $M = 10^{10} M_{\odot}$, $|B| \sim 2.35 \times 10^{-8}$ Gauss \cite{5}. Here, $M_{\odot}$ is the solar mass.
When $\hat{q}$ and $B$ have the same sign, it can be seen from the upper left panel of Fig. \ref{fig:6} that the energy extraction efficiency easily exceeds $100\%$, and remains high even in regions where the decay radius extends beyond the ergosphere. For the same decay radius, when $r_x < 2$, the larger $h$ is, the higher the energy extraction efficiency; when $r_x > 2$, the larger $h$ is, the lower the energy extraction efficiency, which is again consistent with the theoretical derivation. From the upper right panel of Fig. \ref{fig:6}, it can be seen that the farther the decay radius is from the event horizon, the lower the efficiency, and the larger $h$ is, the lower the maximum efficiency. When $\hat{q}$ and $B$ have opposite signs, it can be seen from the lower panel of Fig. \ref{fig:6} that the efficiency is negative and lacks physical meaning. In addition, it can be observed from the figure that when hair $h$ is present, the decay radius stops increasing beyond a certain point; this is because larger decay radii no longer satisfy the condition $0 < h \le 2/\ln(r_x/2)$, a situation that does not occur for the Kerr black hole.

On the other hand, the larger $|B|$ is, the greater the order of magnitude of $\eta$.
\begin{figure}[!h]
  \centering
  \begin{subfigure}{0.45\textwidth}
    \centering
    \includegraphics[width=\linewidth]{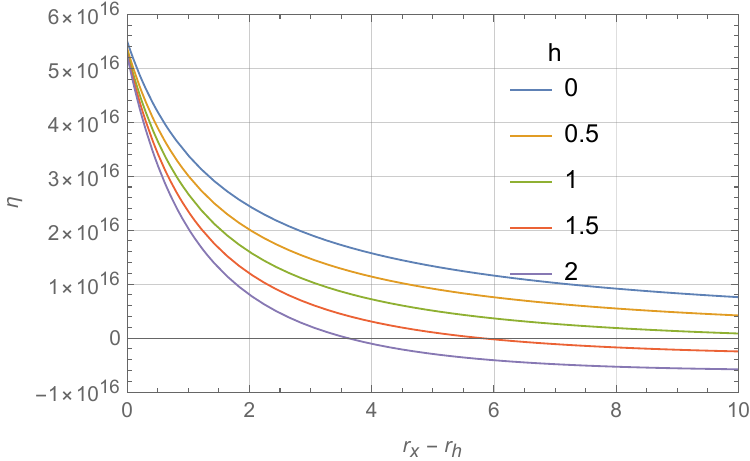} 
  \end{subfigure}
  \begin{subfigure}{0.45\textwidth}
    \centering
    \includegraphics[width=\linewidth]{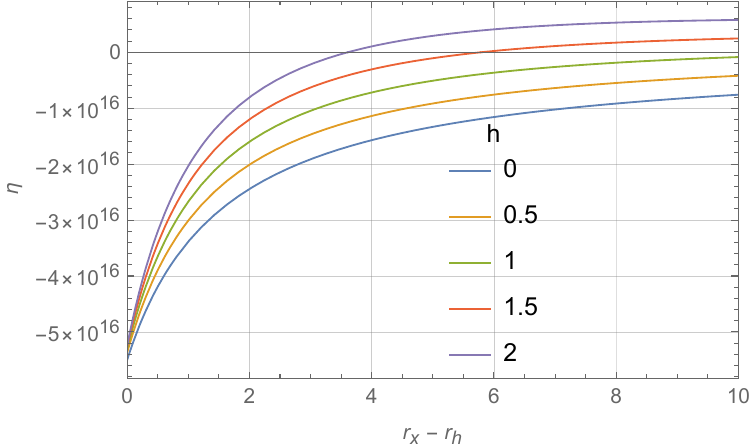} 
  \end{subfigure}
  \caption{$a=0.8$, the variation of energy extraction efficiency with the decay radius for different values of $h$. Left panel: $B=-0.1$, right panel: $B=0.1$.}
\label{fig:7}
\end{figure}
It can be seen from Fig. \ref{fig:7} that for $|B|=0.1$, the energy extraction efficiency reaches the order of $10^{18}\%$. Interestingly, when $\hat{q}$ and $B$ have opposite signs, as shown in the right panel, the efficiency of the black hole with hair becomes positive at high decay radius and remains very high, on the order of $10^{17}\%$, which is not observed for the Kerr black hole.

\section{Conclusion}
In this paper, we have studied the magnetic Penrose process for a rotating black hole in magnetized Horndeski gravity. Compared with the Kerr spacetime, the rotating black hole in Horndeski gravity possesses an additional hair parameter $h$, whose presence significantly alters the ergosphere region. As shown in Fig. \ref{fig:1}, the ergosphere region shrinks as the hair parameter increases.

The negative-energy region is crucial for energy extraction. In the absence of a magnetic field, the negative energy region lies inside the ergosphere, while in the presence of a magnetic field, the situation is quite different. For $\bar{q}B>0$, only a small $\bar{q}B$ is needed for the negative energy region to extend beyond the ergosphere at certain angles. As $\bar{q}B$ increases, the negative energy region continues to expand until it completely extends beyond the ergosphere. However, for $\bar{q}B<0$, the negative energy region remains inside the ergosphere and shrinks as $|\bar{q}B|$ increases, until it disappears. Nevertheless, regardless of the presence or absence of a magnetic field, the negative energy region decreases as the hair parameter increases.

Subsequently, we performed energy extraction via the magnetic Penrose process in the equatorial plane and obtained an expression for the energy extraction efficiency, namely equation \eqref{21} (which is equivalent to equation \eqref{19}). We conducted a detailed theoretical analysis of the relationship between the efficiency expression and the hair parameter. The results show that in the case $\hat{q} B \geq 0$, if the decay radius $r_x > 2$, the efficiency decreases as $h$ increases; if $r_x < 2$, the efficiency increases as $h$ increases; if $r_x = 2$, the efficiency is independent of $h$. However, when $\hat{q}B < 0$, except for the special case $r_x = 2$ where the efficiency is independent of $h$, the variation of efficiency with $h$ depends on the specific values of $r_x$ and $\hat{q}B$, and may exhibit either monotonic decrease or an initial increase followed by a decrease.

Finally, we investigated the influence of the hair parameter on the efficiency through numerical analysis. The numerical simulations yielded results consistent with the theoretical analysis. For the case without a magnetic field, the efficiency is negative and meaningless when $r_x > 2$, and such cases are excluded. For the same decay radius, the larger $h$ is, the higher the efficiency. Furthermore, the farther the decay radius is from the event horizon, the lower the efficiency. The larger $h$ is, the lower the maximum efficiency. For the case with a magnetic field, without loss of generality, we examined a specific process, namely the $\beta$-decay of a neutron into an electron and a proton. When $\hat{q} B > 0$, the energy extraction efficiency easily exceeds $100\%$, and remains high even in regions where the decay radius extends beyond the ergosphere. For the same decay radius, when $r_x < 2$, the larger $h$ is, the higher the energy extraction efficiency; when $r_x > 2$, the larger $h$ is, the lower the energy extraction efficiency, which again agrees with the theoretical derivation. The farther the decay radius is from the event horizon, the lower the efficiency, and the larger $h$ is, the lower the maximum efficiency. When $\hat{q} B < 0$, in the case of a small magnetic field, the efficiency is negative and lacks physical meaning; while in the case of a large magnetic field, such as $|B|=0.1$, the efficiency of the black hole with hair becomes positive at high decay radius and remains very high, on the order of $10^{17}\%$, whereas the efficiency of the Kerr black hole remains negative.

\noindent {\bf Acknowledgments}

\noindent
This work is supported by the National Natural Science Foundation of China (Grants Nos.
12375043, 12575069 ), and Chongqing Normal University Fund Project (Grants No. 26XLB001).


\begin{thebibliography}{99}
\bibitem{10}
R.~Penrose,
``Gravitational collapse: The role of general relativity,''
Riv. Nuovo Cim. \textbf{1} (1969), 252-276
\bibitem{8}
R.~Penrose and R.~M.~Floyd,
``Extraction of rotational energy from a black hole,''
Nature \textbf{229} (1971), 177-179
\bibitem{11}
J.~M.~Bardeen, W.~H.~Press and S.~A.~Teukolsky,
``Rotating black holes: Locally nonrotating frames, energy extraction, and scalar synchrotron radiation,''
Astrophys. J. \textbf{178} (1972), 347
\bibitem{12}
R.~M.~Wald,
``Energy Limits on the Penrose Process,''
Astrophys. J. \textbf{191} (1974), 231
\bibitem{13}
S.~V.~Dhurandhar,
``Energy-extraction processes from a kerr black hole immersed in a magnetic field. i. negative-energy states,''
Phys. Rev. D \textbf{29} (1984) no.12, 2712-2720
\bibitem{14}
S.~V.~Dhurandhar,
``Energy-extraction processes from a kerr black hole immersed in a magnetic field. ii. the formalism,''
Phys. Rev. D \textbf{30} (1984) no.8, 1625-1631
\bibitem{6}
S.~Parthasarathy, S.~M.~Wagh, S.~V.~Dhurandhar and N.~Dadhich, ``High efficiency of the penrose process of energy extraction from rotating black holes immersed in electromagnetic fields,''
Astrophys. J. \textbf{307} (1986), 38
\bibitem{15}
S.~M.~Wagh, S.~V.~Dhurandhar and N.~Dadhich,
``Revival of penrose process for astrophysical applications,''
Atrophys. J. \textbf{301} (1986), 1018
\bibitem{16}
M.~Bhat, S.~Dhurandhar and N.~Dadhich,
``Energetics of the Kerr-Newman black hole by the penrose process,''
J. Astrophys. Astron. \textbf{6} (1985) no.2, 85-100
\bibitem{17}
S.~M.~Wagh and N.~Dadhich,
``The energetics of black holes in electromagnetic fields by the penrose process,''
Phys. Rept. \textbf{183} (1989) no.4, 137-192
\bibitem{18}
N.~Dadhich, A.~Tursunov, B.~Ahmedov and Z.~Stuchl{\'\i}k,
``The distinguishing signature of Magnetic Penrose Process,''
Mon. Not. Roy. Astron. Soc. \textbf{478} (2018) no.1, L89-L94
\bibitem{19}
R.~M.~Crocker, D.~Jones, F.~Melia, J.~Ott and R.~J.~Protheroe,
``A lower limit of 50 microgauss for the magnetic field near the Galactic Centre,''
Nature \textbf{468} (2010), 65
\bibitem{20}
S.~A.~Olausen and V.~M.~Kaspi,
``The McGill Magnetar Catalog,''
Astrophys. J. Suppl. \textbf{212} (2014), 6
\bibitem{21}
K.~Mori, E.~V.~Gotthelf, S.~Zhang, H.~An, F.~K.~Baganoff, N.~M.~Barriere, A.~Beloborodov, S.~E.~Boggs, F.~E.~Christensen and W.~W.~Craig, \textit{et al.}
``NuSTAR discovery of a 3.76-second transient magnetar near Sagittarius A*,''
Astrophys. J. Lett. \textbf{770} (2013), L23
\bibitem{22}
J.~A.~Kennea, D.~N.~Burrows, C.~Kouveliotou, D.~M.~Palmer, E.~Gogus, Y.~Kaneko, P.~A.~Evans, N.~Degenaar, M.~Reynolds and J.~M.~Miller, \textit{et al.}
``Swift Discovery of a New Soft Gamma Repeater, SGR J1745-29, near Sagittarius A*,''
Astrophys. J. Lett. \textbf{770} (2013), L24
\bibitem{23}
R.~P.~Eatough, H.~Falcke, R.~Karuppusamy, K.~J.~Lee, D.~J.~Champion, E.~F.~Keane, G.~Desvignes, D.~H.~F.~M.~Schnitzeler, L.~G.~Spitler and M.~Kramer, \textit{et al.}
``A strong magnetic field around the supermassive black hole at the centre of the Galaxy,''
Nature \textbf{501} (2013), 391-394
\bibitem{24}
R.~Abuter, A.~Amorim, M.~Baub{\"o}ck, J.~P.~Berger, H.~Bonnet, W.~Brandner, Y.~Cl{\'e}net, V.~Coud{\'e} du Foresto, P.~T.~de Zeeuw and C.~Deen, \textit{et al.}
``Detection of orbital motions near the last stable circular orbit of the massive black hole SgrA*,''
Astron. Astrophys. \textbf{618} (2018), L10
\bibitem{25}
K.~Akiyama \textit{et al.} [Event Horizon Telescope],
``First M87 Event Horizon Telescope Results. VIII. Magnetic Field Structure near The Event Horizon,''
Astrophys. J. Lett. \textbf{910} (2021) no.1, L13
\bibitem{26}
K.~Akiyama \textit{et al.} [Event Horizon Telescope],
``First Sagittarius A* Event Horizon Telescope Results. VII. Polarization of the Ring,''
Astrophys. J. Lett. \textbf{964} (2024) no.2, L25
\bibitem{27}
K.~Akiyama \textit{et al.} [Event Horizon Telescope],
``First Sagittarius A* Event Horizon Telescope Results. VIII. Physical Interpretation of the Polarized Ring,''
Astrophys. J. Lett. \textbf{964} (2024) no.2, L26
\bibitem{28}
A.~Tursunov, Z.~Stuchl{\'\i}k, M.~Kolo{\v{s}}, N.~Dadhich and B.~Ahmedov,
``Supermassive Black Holes as Possible Sources of Ultrahigh-energy Cosmic Rays,''
Astrophys. J. \textbf{895} (2020) no.1, 14
\bibitem{29}
A.~Tursunov, M.~Kolo{\v{s}} and Z.~Stuchl{\'\i}k,
``Constraints on Cosmic Ray Acceleration Capabilities of Black Holes in X-ray Binaries and Active Galactic Nuclei,''
Symmetry \textbf{14} (2022) no.3, 482
\bibitem{30}
K.~Greisen,
``End to the cosmic ray spectrum?,''
Phys. Rev. Lett. \textbf{16} (1966), 748-750
\bibitem{31}
G.~T.~Zatsepin and V.~A.~Kuzmin,
``Upper limit of the spectrum of cosmic rays,''
JETP Lett. \textbf{4} (1966), 78-80
\bibitem{32}
A.~Aab \textit{et al.} [Pierre Auger],
``Observation of a Large-scale Anisotropy in the Arrival Directions of Cosmic Rays above $8 \times 10^{18}$ eV,''
Science \textbf{357} (2017) no.6537, 1266-1270
\bibitem{33}
A.~Aab \textit{et al.} [Pierre Auger],
``An Indication of anisotropy in arrival directions of ultra-high-energy cosmic rays through comparison to the flux pattern of extragalactic gamma-ray sources,''
Astrophys. J. Lett. \textbf{853} (2018) no.2, L29
\bibitem{34}
Z.~Stuchl{\'\i}k and M.~Kolo{\v{s}},
``Acceleration of the charged particles due to chaotic scattering in the combined black hole gravitational field and asymptotically uniform magnetic field,''
Eur. Phys. J. C \textbf{76} (2016) no.1, 32
\bibitem{35}
S.~Shaymatov, N.~Dadhich and A.~Tursunov,
``Energetics of Buchdahl stars and the magnetic Penrose process,''
Eur. Phys. J. C \textbf{84} (2024) no.10, 1015
\bibitem{36}
S.~Shaymatov, P.~Sheoran, R.~Becerril, U.~Nucamendi and B.~Ahmedov,
``Efficiency of Penrose process in spacetime of axially symmetric magnetized Reissner-Nordstr{\"o}m black hole,''
Phys. Rev. D \textbf{106} (2022) no.2, 024039
\bibitem{37}
C.~Chakraborty, P.~Patil and G.~Akash,
``Magnetic Penrose process in the magnetized Kerr spacetime,''
Phys. Rev. D \textbf{109} (2024) no.6, 064062
\bibitem{38}
T.~Xamidov, S.~Shaymatov, P.~Sheoran and B.~Ahmedov,
``Astrophysical insights into magnetic Penrose process around parameterized Konoplya{\textendash}Rezzolla{\textendash}Zhidenko black hole,''
Eur. Phys. J. C \textbf{84} (2024) no.12, 1300
\bibitem{39}
S.~Shaymatov,
``Efficiency of magnetic Penrose process in higher dimensional Myers-Perry black hole spacetimes,''
Phys. Rev. D \textbf{110} (2024) no.4, 044042
\bibitem{40}
D.~P.~Viththani, T.~Bhanja, V.~Patel and P.~S.~Joshi,
``Magnetic Penrose process and Kerr black hole mimickers,''
Phys. Rev. D \textbf{110} (2024) no.12, 123035
\bibitem{41}
V.~Patel, K.~Acharya, P.~Bambhaniya and P.~S.~Joshi,
``Energy extraction from Janis-Newman-Winicour naked singularity,''
Phys. Rev. D \textbf{107} (2023) no.6, 064036
\bibitem{5}
S.~J.~Zhang,
``Penrose process in magnetized non-Kerr rotating spacetime with anomalous quadrupole moment,''
JCAP \textbf{12} (2025), 004
\bibitem{42}
M.~Mirkhaydarov, T.~Xamidov, P.~Sheoran, S.~Shaymatov and H.~Nandan,
``Non-Monotonic Enhancement of the Magnetic Penrose Process in Kerr-Bertotti-Robinson Spacetime and its Implication for Electron Acceleration,''
[arXiv:2601.09919 [gr-qc]].
\bibitem{43}
G.~W.~Horndeski,
``Second-order scalar-tensor field equations in a four-dimensional space,''
Int. J. Theor. Phys. \textbf{10} (1974), 363-384
\bibitem{44}
C.~Deffayet, G.~Esposito-Farese and A.~Vikman,
``Covariant Galileon,''
Phys. Rev. D \textbf{79} (2009), 084003
\bibitem{45}
A.~Nicolis, R.~Rattazzi and E.~Trincherini,
``The Galileon as a local modification of gravity,''
Phys. Rev. D \textbf{79} (2009), 064036
\bibitem{46}
T.~Kobayashi, M.~Yamaguchi and J.~Yokoyama,
``Generalized G-inflation: Inflation with the most general second-order field equations,''
Prog. Theor. Phys. \textbf{126} (2011), 511-529
\bibitem{47}
E.~Babichev, C.~Charmousis and A.~Leh{\'e}bel,
``Black holes and stars in Horndeski theory,''
Class. Quant. Grav. \textbf{33} (2016) no.15, 154002
\bibitem{48}
Y.~P.~Hu, H.~A.~Zeng, Z.~M.~Jiang and H.~Zhang,
``P-V criticality in the extended phase space of black holes in Einstein-Horndeski gravity,''
Phys. Rev. D \textbf{100} (2019) no.8, 084004
\bibitem{1}
M.~Heydari-Fard, M.~Heydari-Fard and N.~Riazi,
``Thin accretion disk images of rotating hairy Horndeski black holes,''
Astrophys. Space Sci. \textbf{369} (2024) no.9, 96
\bibitem{2}
R.~K.~Walia, S.~D.~Maharaj and S.~G.~Ghosh,
``Rotating Black Holes in Horndeski Gravity: Thermodynamic and Gravitational Lensing,''
Eur. Phys. J. C \textbf{82} (2022), 547
\bibitem{49}
Y.~G.~Miao and Z.~M.~Xu,
``Thermodynamics of Horndeski black holes with non-minimal derivative coupling,''
Eur. Phys. J. C \textbf{76} (2016) no.11, 638
\bibitem{50}
X.~J.~Gao, T.~T.~Sui, X.~X.~Zeng, Y.~S.~An and Y.~P.~Hu,
``Investigating shadow images and rings of the charged Horndeski black hole illuminated by various thin accretions,''
Eur. Phys. J. C \textbf{83} (2023), 1052
\bibitem{51}
T.~Kobayashi,
``Horndeski theory and beyond: a review,''
Rept. Prog. Phys. \textbf{82} (2019) no.8, 086901
\bibitem{52}
A.~Cisterna and C.~Erices,
``Asymptotically locally AdS and flat black holes in the presence of an electric field in the Horndeski scenario,''
Phys. Rev. D \textbf{89} (2014), 084038
\bibitem{53}
X.~J.~Wang, X.~M.~Kuang, Y.~Meng, B.~Wang and J.~P.~Wu,
``Rings and images of Horndeski hairy black hole illuminated by various thin accretions,''
Phys. Rev. D \textbf{107} (2023) no.12, 124052
\bibitem{54}
X.~X.~Zeng, C.~Y.~Yang, M.~I.~Aslam and R.~Saleem,
``Probing Horndeski gravity via Kerr black hole: Insights from thin accretion disks and shadows with EHT observations,''
JHEAp \textbf{51} (2026), 100540


\bibitem{3}
R.~M.~Wald,
``Black hole in a uniform magnetic field,''
Phys. Rev. D \textbf{10} (1974), 1680-1685
\bibitem{4}
A.~N.~Aliev and N.~Ozdemir,
``Motion of charged particles around a rotating black hole in a magnetic field,''
Mon. Not. Roy. Astron. Soc. \textbf{336} (2002), 241-248



\bibitem{7}
A.~Tursunov and N.~Dadhich,
``Fifty years of energy extraction from rotating black hole: revisiting magnetic Penrose process,''
Universe \textbf{5} (2019) no.5, 125

\end{thebibliography}
\end{document}